\documentstyle[12pt]{article}
\def\be{\begin{eqnarray}}
\def\ee{\end{eqnarray}}
\def\nn{\nonumber}
\def\ds{\displaystyle}
\def\H{{\cal H}}
\def\lap{\triangle}
\def\V{V_{C}}
\def\R{{\cal R}}
\def\t{\mbox{\bf t}}
\def\n{\mbox{\bf n}}
\def\b{\mbox{\bf b}}
\def\r{\mbox{r}}
\def\rb{\mbox{\bf r}}
\def\x{\mbox{x}}
\def\y{\mbox{y}}
\def\z{\mbox{z}}
\def\I{\mbox{\bf 1}}
\def\w{w}
\def\hn{{\hat\nabla}}
\def\tr{\mbox{tr}}

\begin{document}
\begin{flushright}
{\large{\sl UPRF-94-408}}
\end{flushright}
\vskip1.5cm
\begin{center}
{\LARGE A Complete Perturbative Expansion}\\
\medskip
{\LARGE for Constrained Quantum Dynamics}\\
\vskip0.8cm
{\Large P.\ Maraner}\\
\smallskip
{\large\sl Dipartimento di Fisica, Universit\`a di Parma,}\\
\smallskip
{\large\sl and INFN, Gruppo collegato di Parma,}\\
\smallskip
{\large\sl  Viale delle Scienze, 43100 Parma, Italy}
\vskip0.6cm
{\large August 1994}
\end{center}
\vskip2cm
\begin{abstract}
A complete perturbative expansion for the Hamiltonian describing the motion of
a quantomechanical system constrained to move on an arbitrary submanifold of
its
configuration space $R^n$ is obtained.
\end{abstract}

\thispagestyle{empty}
\newpage

\section{Introduction}

The quantomechanical description of constrained systems is extremely important
in physics and since the early days of quantum mechanics several techniques
have been developed to deal with this matter. A  fundamental contribution has
been given by Dirac \cite{Dir}. His idea, geometrical in nature, consists
in removing the redundant degrees of freedom by the construction of a
consistent
hamiltonian formalism for the constrained classical theory and proceeding then
to its quantization. Other noteworthy approach have been developed by
Schwinger,
Peierls by using variational arguments and by De Witt, Faddeev, Popov
by means of lagrangian formalism and of path integral techniques. The common
feature to all this methods, to which we will refer as {\sl formal methods},
is that {\sl the reduction of the dynamics to the constraint
surface is performed before the quantization of the system}.
Although in many cases this is the only way to proceed,
it often introduces non physical ambiguities
causing some pure quantomechanical effects to be ignored.

Let us consider for example a particle constrained to move on an arbitrary
surface $\Sigma$ embedded in the three-dimensional euclidean space $R^3$.
The reduction of the classical theory is straightforward.
Introducing $x^1,x^2$ coordinates parametrizing the surface
and denoting by $g_{\mu\nu}$, $\mu,\nu=1,2$,
the metric induced  on $\Sigma$ from $R^3$, the system is described by the
lagrangian ${\cal L}={1\over 2}g_{\mu\nu}{\dot x^\mu}{\dot x^\nu}$.
Defining the generalized momenta
$p_\mu=\partial {\cal L}/ \partial {\dot x^\mu}$
we obtain the Hamiltonian
${\cal H}={1\over 2}g^{\mu\nu}p_\mu p_\nu$, $g^{\mu\nu}$ denoting the
inverse of the metric. The quantization of $\H$ contains ordering
ambiguities which are not completely removed by the required covariance of the
theory. As first observed by De Witt \cite{DeW}, in constructing  the
Hamiltonian operator we are free to add to minus one half the laplacian $\lap$
a term proportional to the scalar curvature $R$ of the surface,
\be
\H=-{1\over 2}\lap + \alpha R.
\ee
Different quantization schemes produces different values for the constant
$\alpha$. $\alpha$ may not be unambiguously determined, depending
essentially on an ordering choice. On the other hand the reduction of the
motion of a particle to a surface is by no means an academic problem. Devices
producing the confinement of electrons on a plane are widely studied in
physics, for example in the Quantum Hall Effect's context \cite{Pra},
and we may think to use the same techniques to constraint a particle
on an arbitrary surface $\Sigma$. How to determine then the constant $\alpha$?
 In addressing the solution of the problem it is convenient to give up the
formal treatment of the constraint, thinking instead at the physical mechanism
producing the confinement of the particle to the surface. The analysis of
devices used in the Quantum Hall Effect stuff, suggests the confinement to be
produced by  a potential presenting a deep minimum in correspondence of the
constraint surface and depending only on the coordinate normal to it
\cite{Pra,MD,Mar}.
Moving along this line H. Jensen and H. Koppe \cite{JK} have first given a
realistic description of the motion of a particle on a surface embedded in
$R^3$. In accordance with Heisenberg's principle the confinement causes the
particle to fluctuate very strongly in the direction normal to the surface so
that the spectrum of the system is described in first approximation by that of
the confining potential. In correspondence of each level the effective
Hamiltonian describing the motion along the surface may then be unambiguously
obtained as (see Ref.\cite{JK} and the discussion below)
\be
\H=-{1\over 2}\lap + {1\over 4} R - {1\over 8}\xi^2,
\label{iHS}
\ee
where $\xi$ is the extrinsic mean curvature of the surface $\Sigma$. The
analysis have to be completed by considering the interactions between the
degrees of freedom normal to the surface and those parallel to it.
The case of a wire embedded in $R^3$
has also been considered by many authors \cite{W}.

{}From this simple example we learn that the operation of reducing the dynamics
and quantizing a constrained system do not in general commute and that
performing the former
before the latter may produce the appearance of unphysical ambiguities and the
neglecting of contributions connected to the extrinsic geometrical properties
of the constraint surface.

 In this paper we present a complete perturbative description of a system
constrained to move on a submanifold of its configuration space $R^n$ by a
confining potential $V_C$. Generalizing the example  of the surface embedded in
$R^3$ we require $V_C$ to satisfy  two very general conditions
\begin{description}
\item{C1)} $V_C$ presents a deep minimum in correspondence of the constraint
surface,
\item{C2)} $V_C$ depends only on coordinates normal to the constraint surface.
\end{description}

In our scheme {\sl the reduction of the dynamics to the constraint surface is
performed after the quantization of the system}. Adapting coordinates to
the constraint in section 2, a complete perturbative expansion for the
Hamiltonian describing the motion of the system is obtained in section 3.
In accordance with
Heisenberg principle the zero order term of the  expansion
takes into account the fluctuations of the
system in the directions normal to the constraint surface. The first order
term, already discussed in Refs.\cite{MD,Mar,FO},  describes the effective
constrained dynamics while the rest of the perturbative expansion describes
the interactions between normal and effective degrees of freedom. We want to
point out that the constrained quantum dynamics is characterized by the whole
expansion, the effective dynamics on the constraint surface representing only
the leading term. Contrary to what happens in the classical description,
the explicit form of the potential  realizing the constraint  leaves
traces in the effective dynamics and in the spectrum of the system
and therefore may not be neglected.
This is illustrated by two example in sections 5 and 6.
Section 7 contains our conclusions.

\section{Geometrical Preliminary}

In the sequel we identify the
constraint surface with a smooth $m$-dimensional submanifold $M$ of the
configuration space $R^n$. Denoting by $\Phi:M\rightarrow R^n$ the embedding of
$M$ in $R^n$ and by $\n^1(x),\n^2(x),...,\n^{n-m}(x)$, a smooth assignment of
$(n-m)$ orthonormal vectors normal to $M$ in every point $x\in M$, an
{\sl adapted coordinates frame} may be introduced by using coordinates
$x^\mu,\mu=1,...,m$, on $M$, plus the distances $y^i,i=1,...,n-m$,
along the geodetics leaving $M$ with speed $\n^i$.
In a  ``sufficently close'' neighbourhood  of $M$ the frame
$\{x^\mu,y^i;\mu=1,...,m,i=1,...,n-m\}$ is well defined
and its relation with the Cartesian coordinates $\mbox{\bf r}=(r^1,...,r^n)$
of $R^n$ is given by
\be
\mbox{\bf r}=\Phi(x^{\mu})+y^i\n^i(x^{\mu}).
\ee

It is important to realize that the embedding of $M$ in $R^n$ is completely
characterized by the assignment of some tensorial quantities on $M$
\cite{Spi}. In an adapted coordinates frame these may be easily constructed
as follows
\be
&g_{\mu\nu}=\t_\mu\cdot\t_\nu &
     \mbox{{\sl induced metric (first fundamental form)}}\nn\\
&\alpha^i_{\mu\nu} = \n^i\cdot\partial_\nu\t_\mu &
                                       \mbox{{\sl second fundamental
form}}\nn\\
&A_\mu^{ij} =\n^i\cdot\partial_\mu \n^j&\mbox{{\sl normal fundamental form}}\nn
\ee
where $\t_\mu=\partial_\mu\Phi$ denote the tangent vectors to $M$ associated
with the chosen coordinate frame and the dot the standard scalar product in
$R^n$.

The choice of an adapted coordinates frame is obviously not unique. An
arbitrary coordinates transformation on $M$ as well as a point dependent
rotation of the normal vectors $\n^i(x)$ transforms an adapted coordinates
frame into an adapted coordinates frame. Whereas
varying the choice of the coordinates $x^\mu$ causes $g_{\mu\nu}$,
$\alpha_{\mu\nu}^{i}$ and $A_{\mu}^{ij}$ to transform as tensors of $M$,
the variation of normal vectors $\n^i(x)$ by a rotation $R^{kl}(x)$ makes
$\alpha_{\mu\nu}^{i}$ to transform as a $SO(n-m)$ vector but $A_{\mu}^{ij}$
as a $SO(n-m)$ gauge connection
\be
A_\mu^{ij}\longrightarrow \R^{ik}A_\mu^{kl}\R^{jl}+\R^{ik}\partial_\mu\R^{jk}.
\ee
 The normal fundamental form $A_{\mu}^{ij}$ actually represents the connection
induced by $R^n$ on the normal bundle of $M$, $TM^{\perp}$.

The metric $G_{IJ}$, $I,J=1,...,n$
of $R^n$ in the adapted coordinates frame $\{x^\mu,y^i\}$ writes
\be
G_{IJ} = \pmatrix{
    \gamma_{\mu\nu}+y^k y^l A_{\mu}^{kh} A_{\nu}^{lh}&
       y^k A_{\mu}^{jk} \cr
       y^k A_{\nu}^{ik}  & \delta^{ij}}
\label{G}
\ee
where, introducing the matrix $\eta$ by
$\eta^\mu_\nu=y^ig^{\mu\rho}\alpha^i_{\rho\nu}$, the matrix $\gamma$ may
be written as
\be
\gamma_{\mu\nu}=g_{\mu\rho}(\I-\eta)^\rho_\sigma(\I-\eta)^\sigma_\nu
\label{gamma}
\ee
The determinant $|G|$ of $G_{IJ}$ coincides with that of the matrix $\gamma$,
$|\gamma|$, and the inverse of the metric tensor may be calculated as
\be
G^{IJ} = \pmatrix{
    \gamma^{\mu\nu}&
       \gamma^{\mu\rho} y^k A_{\rho}^{kj} \cr
       \gamma^{\nu\rho} y^k A_{\rho}^{ki}  &
       \delta^{ij}+ y^k y^l A_{\rho}^{ik} A_{\sigma}^{jl}
                          \gamma^{\rho\sigma}}.
\label{Gi}
\ee

\section{The Perturbative Expansion}

We come now to the dynamical aspect of the problem. The quantization of
the system is performed very easily and unambiguously in $R^n$, before
considering the constraint. In a cartesian coordinates frame
$\{r^I;I=1,...,n\}$, dynamics is described by
\be
\H=-{1\over 2}\partial_I\partial_I+\V,
\label{Ha}
\ee
where $\V$ is the potential realizing the constraint. $\H$  acts on
wavefunctions $\psi\in{\cal L}^2(R^n)$ normalized with the condition
\be
\int |\psi|^2 dr^n = 1.
\label{nc1}
\ee
In an {\sl adapted coordinates frame} $\{x^\mu,y^i;\mu=1,...,m,i=m+1,...,n\}$
Hamiltonian (\ref{Ha}) takes the form
\be
{\cal H}=-{1\over 2|G|^{1/2}}\partial_I G^{IJ}|G|^{1/2}\partial_J+\V,
\label{Hb}
\ee
and the normalization condition (\ref{nc1}) transforms to
\be
\int |\psi|^2 |G|^{1/2} dx^m dy^{(n-m)} =1.
\label{nc2}
\ee
Since we are looking for an effective dynamics on the submanifold $M$ we find
it natural to perform a similitude transformation in such a way that wave
functions are correctly normalized in ${\cal L}^2(M)$ instead of
${\cal L}^2(R^n)$. The aim is achieved by
\be
\begin{array}{rcl}
\psi  &\rightarrow&  \ds{|g|^{1/4}\over |G|^{1/4}}\psi \cr
\H    &\rightarrow&  \ds{|g|^{1/4}\over |G|^{1/4}}\H{|G|^{1/4}\over |g|^{1/4}}
\end{array}
\ee
where $|g|$ denotes the determinant of the metric $g_{\mu\nu}$ induced on $M$.
Considering the explicit form (\ref{Gi}) of the inverse  metric $G^{IJ}$
and introducing $\hat\partial_\mu=\partial_\mu+iA_\mu^{ij}L_{ij}/2$, where
$L_{ij}=-i(y^i\partial_j-y^j\partial_i)$ are the $SO(n-m)$ generators,
Hamiltonian (\ref{Hb}) takes the quite complicated form
\be
\H=-{1\over 2|\gamma|^{1/4}}
    \partial_i|\gamma|^{1/2}\partial_i{1\over|\gamma|^{1/4}}
   -{1\over 2|g|^{1/4}|\gamma|^{1/4}}
    \hat\partial_\mu \gamma^{\mu\nu} |\gamma|^{1/2}\hat\partial_\nu
    {|g|^{1/4}\over|\gamma|^{1/4}}+\V.
\label{Hc}
\ee

At this point, and only at this point, the constraint is imposed by
considering conditions $C1$ and $C2$. Condition $C1$ assures that
the potential $\V$ may be replaced by its power expansion around
the minimum ${\vec y}=0$.  Condition $C2$ states that there exist an
adapted coordinates frame in which $\V$ only depends
on the normal coordinates ${\vec y}$.
 Without loss of generality the constant term of the expansion may be neglected
and the quadratic term diagonalized by means of a point independent rotation in
the normal space
\be
\V({\vec y})={1\over 2\epsilon^2}{\omega^i}^2{y^i}^2+
             a_{ijk}y^iy^jy^k+b_{ijkl}y^iy^jy^ky^l+...\ .
\label{cp}
\ee
The scale of the proper frequencies $\omega^i$ has been readsorbed in the
adimensional parameter $\epsilon^{-1}$. The smaller $\epsilon$ the deeper is
the minimum of $\V$ and the more the system is squeezed on the constraint
surface.

 $\epsilon$ appears as a natural perturbative parameter in the theory and,
rescaled the normal coordinates by ${\vec y}\rightarrow
\epsilon^{1/2}{\vec y}$,
a perturbative theory may be set up by expanding  Hamiltonian (\ref{Hc})
in powers of $\epsilon$
\be
\begin{array}{rcl}
\epsilon{\cal H}&=&H^{(0)}+\epsilon       H^{(1)}
                        +\epsilon^{3/2} H^{(3/2)}
                        +\epsilon^{2}   H^{(2)}   +...\  + \\
                & &+\epsilon^{5/2}a_{ijk}y^iy^jy^k+
                   \epsilon^{3}b_{ijkl}y^iy^jy^ky^l+...\ .
\end{array}
\label{Hexp}
\ee
The constants $a_{ijk}$, $b_{ijkl}$, ... appearing in the expansion of the
confining potential are such that the second, the third and further terms
of the right hand side of (\ref{cp}) are small compared to the first term and
in this sense are $\epsilon$-dependent.
In practical application they appear in the perturbative
expansion as independent parameter so that, for example, $\epsilon^{5/2}a_{ijk}
y^iy^jy^k$ is not in general of order $\epsilon^{5/2}$, its magnitude depending
on the explicit form of the potential $\V$. The zero and first order terms of
expansion (\ref{Hexp}) has been extensively discussed in Ref.\cite{MD}.

In accordance with Heisenberg principle the zero order dynamics depends only
on normal degrees of freedom,
\be
H^{(0)}={1\over 2}\left(-\partial_i\partial_i+{\omega^i}^2{y^i}^2\right),
\label{H0}
\ee
describing a system of $(n-m)$ uncoupled harmonic oscillators with frequencies
$\omega^{m+1},...\omega^{n}$.

More surprising results follow from the analysis of the first order term,
\be
  H^{(1)}=-{1\over 2 g^{1/2}}
   \left( \partial_\mu+{i\over 2}A_\mu^{ij}L_{ij}\right)
   g^{\mu\nu}g^{1/2}
   \left( \partial_\nu+{i\over 2}A_\nu^{kl}L_{kl}\right)
   +Q^{(1)}(x),
\ee
where the potential $Q^{(1)}$ may be expressed in terms of the intrinsic
scalar curvature $R$ and the extrinsic mean curvature $\xi$ as
\be
Q^{(1)}(x)={1\over 4} R(x) - {m^2\over 8} \xi^2(x).
\label{iQ}
\ee
Aside from the potential term $Q^{(1)}$, $H^{(1)}$ is proportional to the
Laplace operator on $M$ coupled to the motion in normal directions by means of
the minimal interaction with the gauge field $A_\mu^{ij}L_{ij}/2$.
 It is therefore resonable to expect that in a perturbative picture
$H^{(1)}$ describes the effective dynamics on $M$. This has actually been found
in  Ref.\cite{MD}. The surprising result, unexpected and
unrecoverable by means of a formal treatment of constraints, is that the
effective dynamics is coupled with  gauge fields and quantum
potentials induced by the intrinsic and extrinsic geometrical properties of the
constraint surface. The physical relevance of such a geometry-induced
dynamical structure has been recently discussed in Ref.\cite{Mar}, showing how
this phenomenon is observable in the effective rotational motion of some
simple polyatomic molecules.

Since we are interested in a realistic description of a constrained microscopic
system we never consider the limit $\epsilon\rightarrow 0$. $\epsilon$ is a
small but finite parameter, its magnitude depending on the characteristics of
the system under consideration. It is therefore very important to know the
explicit expression of further terms of expansion (\ref{Hexp}) in order to
predict the spectrum of the system with an adequate precision.

 To evaluate the explicit expression of the generic term of the expansion
(\ref{Hexp}) we start by observing that the first and second terms of
Hamiltonian (\ref{Hc}) may be rewritten solely in terms of $\gamma^{\mu\nu}$
and $\ln|\I-\eta|$ as
\be
-{1\over 2|\gamma|^{1/4}}\partial_i|\gamma|^{1/2}
 \partial_i{1\over|\gamma|^{1/4}}=
-{1\over 2}\partial_i\partial_i+\nn\\
+{1\over 4}(\partial_i\partial_i\ln|\I-\eta|)
+{1\over 8}(\partial_i\ln|\I-\eta|)(\partial_i\ln|\I-\eta|),\nn
\ee
and
\be
-{1\over 2|g|^{1/4}|\gamma|^{1/4}}
\hat\partial_\mu \gamma^{\mu\nu} |\gamma|^{1/2}\hat\partial_\nu
{|g|^{1/4}\over|\gamma|^{1/4}}=-{1\over 2}\hn_\mu\gamma^{\mu\nu}
\hn_\nu +\nn\\
+{1\over 4}\left(\hn_\mu\gamma^{\mu\nu}\hn_\nu\ln|\I-\eta|\right)
 +{1\over 8}\gamma^{\mu\nu} \left(\hn_\mu\ln|\I-\eta|\right)
                  \left(\hn_\nu\ln|\I-\eta|\right),\nn
\ee
where, denoting by $\nabla_\mu$ the covariant derivative associated with the
connection induced on $M$,
\be
\hn_\mu=\nabla_\mu+{i\over 2}A_\mu^{ij}L_{ij}.
\label{cd}
\ee
It is  very convenient to introduce the matrices
\be
\eta^{\mu\nu}_{(N)}=(N+1)g^{\mu\rho_1}y^{i_1}
\alpha^{i_1}_{\rho_1\sigma_1}g^{\sigma_1\rho_2}...y^{i_N}
\alpha^{i_N}_{\rho_N\sigma_N}g^{\sigma_N\nu},
\ee
$\eta_{(0)}^{\mu\nu}=g^{\mu\nu}$. The expansion in $\epsilon$ of
$\gamma^{\mu\nu}$ and $\ln|\I-\eta|$ may then be computed as
\be
\gamma^{\mu\nu}=
\sum_{N=0}^{\infty}\epsilon^{N/2}\eta^{\mu\nu}_{(N)},
\label{1}
\ee
\be
\ln|\I-\eta|=-\sum_{N=1}^{\infty}{\epsilon^{N/2}\over N}
 \tr\left[\eta^N\right].
\label{2}
\ee
The evaluation of the $N/2$-order term of the perturbative expansion
(\ref{Hexp}) reduces so to a matter of simple algebra  yielding
\be
H^{(N/2)}  = -{1\over 2}\hn_\mu \eta_{(N-2)}^{\mu\nu}\hn_\nu+ Q^{(N/2)},
\label{HN}
\ee
$N\geq 2$, where the potentials $Q^{(N/2)}$ may be written as
\be
Q^{(1)}  = {1\over 8} \tr\left[\partial_i\eta\right]
                       \tr\left[\partial_i\eta\right]
           -{1\over 4} \tr\left[\partial_i\eta\partial_i\eta\right],\nn\\
Q^{(3/2)}= {1\over 4} \tr\left[\partial_i\eta\right]
                       \tr\left[\eta\partial_i\eta\right]
           -{1\over 2} \tr\left[\eta\partial_i\eta\partial_i\eta\right]
           -{1\over 4} \tr\left[\hn_\mu g^{\mu\nu} \hn_\nu \eta \right],\nn
\ee
and for $N\geq 4$
\be
Q^{(N/2)}=\sum_{K=0}^{N-2}
\left\{
 {1\over 8} \tr\left[\eta^K\partial_i\eta\right]
            \tr\left[\eta^{N-K-2}\partial_i\eta\right]
-{1\over 4} \tr\left[\eta^{N-2}\partial_i\eta\partial_i\eta\right]
\right\}+\nn  \\
         - \sum_{K=0}^{N-3} {1\over 4}
            \tr\left[\eta^{N-K-3}\hn_\mu\eta^{\mu\nu}_{(K)}\hn_\nu\eta +
            (N-K-3)\eta^{\mu\nu}_{(K)}
            \eta^{N-K-4}(\hn_\mu\eta)(\hn_\nu\eta)\right]+\nn\\
         +\sum_{K=0}^{N-4} \sum_{L=1}^{N-K-3} {1\over 8}
           \eta^{\mu\nu}_{(K)}
           \tr\left[\eta^{L-1}(\hn_\mu\eta)\right]
           \tr\left[\eta^{N-K-L-3}(\hn_\nu\eta)\right].\nn
\ee

 It is remarkable that the perturbative expansion of the Hamiltonian
of the system is completely written in terms of the induced metric
$g_{\mu\nu}$, the second fundamental form $\alpha_{\mu\nu}^{i}$, its first and
second covariant derivative $\nabla_\rho\alpha_{\mu\nu}^{i}$,
$\nabla_\sigma\nabla_\rho\alpha_{\mu\nu}^{i}$ and the normal fundamental form
$A_{\mu}^{ij}$. $A_{\mu}^{ij}$ appears in the perturbative expansion only by
means of the minimal coupling (\ref{cd}).

\section{Spectrum and Effective Dynamics}

To evaluate the spectrum of the system we proceed now by means of the
standard Raleigh-Schr\"odinger perturbation theory. We identify $H^{(0)}$
with the unperturbed Hamiltonian and the rest of expansion (\ref{Hexp})
with the perturbation ${\cal P}_\epsilon$,
\be
\epsilon\H= H^{(0)} +{\cal P}_\epsilon.
\ee

As stated in the previous section, $H^{(0)}$ represents a system of $(n-m)$
uncoupled harmonic oscillator. We denote by $\chi_{\cal N}({\vec y})$ its
eigenfunctions, having collected the harmonic oscillator quantum numbers
$n_{m+1},...,n_{n}$ in the multiindex ${\cal N}=(n_{m+1},...,n_{n})$. The
corresponding eigenvalues are given by $E^{(0)}=\sum_i \omega^i (n_i+1/2)$.
The spectrum is
degenerate every time  the frequencies $\omega^i$ satisfy linear
conditions in the integer field. The zero order eigenfunctions corresponding to
an energy $E^{(0)}$ are given by
\be
\psi^{(0)}_{\cal N}({\vec x},{\vec y})
= \phi_{\cal N}({\vec x}) \chi_{\cal N}({\vec y})
\ee
and  present an infinite degeneracy  given by the presence of the
arbitrary function of ${\vec x}$, $\phi_{\cal N}({\vec x})$,  beside that
labelled by the multiindex corresponding to the energy $E^{(0)}$.

The first order correction to $E^{(0)}$, $E^{(1)}$, is obtained by
diagonalizing the perturbation on degenerate states, that is by solving the
Schr\"odinger equation
\be
\H^{E^{(0)}}\phi({\vec x})=E^{(1)} \phi({\vec x}),
\label{ise}
\ee
where the Hamiltonian  $\H^{E^{(0)}}$ is obtained by bracketing the order
$\epsilon$ term of ${\cal P}_\epsilon$, $H^{(1)}$, between the harmonic
oscillator states corresponding to $E^{(0)}$ and  $\phi({\vec x})$
is a vector wavefunction having as component the $\phi_{\cal N}({\vec x})$
with energy $E^{(0)}$. The explicit expression of $\H^{E^{(0)}}$ is
\be
\H^{E^{(0)}}=-{1\over 2g^{1/2}}(\I \partial_\mu+iA_\mu)g^{\mu\nu}g^{1/2}
 (\I \partial_\nu+iA_\nu) +Q^{(1)}({\vec x})+{\bar Q}^{(1)}({\vec x})
\label{iH}
\ee
with
\be
&&A_\mu={1\over 2}A_{\mu}^{ij}\langle L_{ij}\rangle,\label{iA} \\
&&{\bar Q}^{(1)}={1\over 8}g^{\mu\nu}A_\mu^{ij}A_\nu^{kl}\left(
\langle L_{ij}L_{kl}\rangle -\langle L_{ij}\rangle \langle L_{kl}\rangle\right)
,\label{Qbar}
\ee
where $\langle L_{ij}\rangle$ and $\langle L_{ij}L_{kl}\rangle$ denote the
matrices obtained by bracketing $L_{ij}$ and $L_{ij}L_{kl}$ between the
harmonic
oscillator states corresponding to $E^{(0)}$ and $\I$ is the identity matrix
with the dimension of the degenerate space.  $\H^{E^{(0)}}$ appears as a free
Hamiltonian  on the constraint surface coupled with the geometry induced gauge
fields (\ref{iA}) and  the potentials (\ref{iQ}) and (\ref{Qbar}). Equation
(\ref{ise}) have therefore to be interpreted as the Schr\"odinger equation
describing the effective dynamics induced on the constraint surface.
Note that for a surface ${\Sigma}$ embedded in the three-dimensional euclidean
space $R^{3}$ the Hamiltonian (\ref{iH}) reduces to (\ref{iHS}).
We do
not comment anymore on this fact referring to \cite{MD,Mar} for details.

 Denoting by ${\cal K}$ the quantum numbers labelling the eigenfunctions of
$\H^{E^{(0)}}$  and supposed the degeneracy to be completely removed, the
eigenvalues of $\epsilon\H$ are evaluated by means of the standard formula
\be
\epsilon{\cal E}_{{\cal N},{\cal K}}= E^{(0)}+
\langle {\cal N},{\cal K}|{\cal P}_\epsilon |{\cal N},{\cal K}\rangle +
\sum_{({\cal N}'{\cal K}')\neq({\cal N},{\cal K})}
{ |\langle {\cal N},{\cal K}|{\cal P}_\epsilon |{\cal N}',{\cal K}'\rangle|^2
\over E^{(0)}-{E^{(0)}}' } + ...\ \ .
\ee
This allows to calculate the spectrum of the system with an arbitrary
accuracy as a power series in the parameter $\epsilon$.

\section{Particle Constrained on a Circle}

As a very simple but nontrivial example we consider a particle constrained to
move on a circle embedded in $R^3$ by an harmonic potential.
This allows  to illustrate some
peculiarities of constrained quantomechanical systems which are systematically
ignored in formal treatments. Let therefore be $c:[0,2\pi R]\rightarrow R^3$,
$c(x)=(R\cos(x/R),R\sin(x/R),0)$ the  embedding map of the circle in the
three-dimensional euclidean space $R^3$.
The curve is parametrized by the arclength
$x$, so that  its tangent, normal and binormal may be immediately evaluated
as $\t(x)=(-\sin(x/R),\cos(x/R),0)$, $\n(x)=(\cos(x/R),\sin(x/R),0)$ and
$\b(x)=(0,0,1)$. Every smooth assignment of an orthonormal basis of the normal
space to $c$ in $x$ may  be obtained by rotating normal and binormal by a
point dependent angle $\w(x)$
\be
\begin{array}{l}
\ds \n^2=\ \cos\w\ \n+\sin\w\ \b,\\
\ds \n^3=- \sin\w\ \n+\cos\w\ \b,
\end{array}
\ee
where $\w(0)=\w(2\pi R)+2\pi z$, $z$ being an integer. The induced metric,
the second fundamental form and the normal fundamental form of the embedding
read
\be
\begin{array}{l}
\ds g_{11}=1,\\
\ds\alpha_{11}^{2}=\ {1\over R}\cos\w,\\
\ds\alpha_{11}^{3}=- {1\over R}\sin\w,\\
\ds A_{1}^{23}= -{\dot \w}.
\end{array}
\ee
$\eta$ is so  written as $\eta={y^2\over R}\cos\w-{y^3\over R}
\sin\w$, whereas the covariant derivative (\ref{cd}) on $c$ reads
${\hat\nabla}_x=\partial_x-i{\dot\w}L_{23}$. The direct calculation shows that
${\hat\nabla}_x\eta=0$, so that the whole perturbative expansion (\ref{Hexp})
may be easily evaluated as
\be
\begin{array}{l}
\ds H^{(0)}={1\over 2}\left(-\partial^2_2
                            +{\omega^2}^2{y^2}^2\right) +
            {1\over 2}\left(-\partial^2_3
                            +{\omega^3}^2{y^3}^2\right), \\
\ds H^{(1)}=-{1\over 2}\left(\partial_x-i{\dot\w}L_{23}\right)^2-
             {1\over 8R^2}, \\
\ds ...,  \\
\ds H^{(N/2)}=(N-1)\left({y^2\over R}\cos\w-{y^3\over R}\sin\w\right)^{N-2}
              H^{(1)},\\
\ds ... \ \ .
\end{array}
\ee
The spectrum of the system may now be calculated by means of perturbation
theory. As in the general case the infinite degeneracy of the zero order states
is removed by solving the Schr\"odinger equation (\ref{ise}) for the effective
dynamics on the constraint surface. In correspondence to the zero order state
labelled by the harmonic oscillator quantum numbers $(n_2,n_3)$ the effective
Hamiltonian on the circle writes
\be
\H^{(n_2,n_3)}=-{1\over 2}\left(\partial_x
               -i{\dot\w}\langle L_{23}\rangle\right)^2+
               {1\over 2}\left(\langle L_{23}^2\rangle-
                              \langle L_{23}\rangle^2\right){\dot\w}^2-
                         {1\over 8R^2},
\ee
where angled brackets denote again  expectation values
between  harmonic oscillator states corresponding to the energy $E^{(0)}=
\omega^2(n_2+1/2)+\omega^3(n_3+1/2)$.

Let us now discuss the physical meaning of the function $\w(x)$. If the
confining potential is symmetric, that is $\omega^2=\omega^3$, it is possible
to choose the harmonic oscillator basis is such a way that $L_{23}$ is
diagonal. The effective potential $(\langle L_{23}^2\rangle-
\langle L_{23}\rangle^2){\dot\w}^2/2$ vanishes then identically and
$\langle L_{23}\rangle{\dot\w}$ appears then as a pure  gauge field in the
theory and may be removed by a different choice of normal coordinates
$y^2,y^3$. The
effective dynamics on the circle and the whole perturbative expansion result
then considerably simplified. On the contrary, if the confining potential is
not symmetric, $\omega^2\neq\omega^3$, $\langle L_{23}\rangle=0$,
$\langle L_{23}^2\rangle\neq 0$ and a different choice of normal coordinates
would cause the confining potential to be $x$-dependent.
The effects produced by $\w(x)$ may therefore not be eliminated.
 Pictorially we may assimilate our model to a particle moving in a ring with
a small ellipsoidal section. The function $\w(x)$ describes then how
the section wraps up when moving along the ring.
If the ring's section reduces to a circle ($\omega^2=\omega^3$)
then does not matter how the wrapping is done
and we are always reconduced to the case
$\w=0$. On the contrary the wrapping produces observable effects when the
ring's section is not circular ($\omega^2\neq\omega^3$).

 The case $\omega^2=\omega^3$ being straightforward we concentrate on
$\omega^2\neq\omega^3$. The effective Hamiltonian describing the dynamics on
the circle reduces then to
\be
\H^{(n_2,n_3)}=-{1\over 2}\partial_x^2+{1\over 2}
\left[\left({\omega^2\over\omega^3}+{\omega^3\over\omega^2}\right)
      \left(n_2+{1\over 2}\right)\left(n_3+{1\over 2}\right)
-{1\over 2}\right]{\dot\w}^2-{1\over 8R^2}.
\ee
Different choices of the {\sl wrapping function} $\w(x)$ produce a completely
different effective dynamics. An arbitrary positive, everywhere finite
smooth potential
may be reproduced by a suitable choice of $\w$.

 The simpler case we may consider is that in which the potential wraps us
uniformly, say $z$ times, $\w(x)=zx/R$. The
effective Schr\"odinger equation on the circle is then immediately solved by
$\phi_{(n_2,n_3),k}(x)= e^{i{k\over R}x}/\sqrt{2\pi R}$, $k$ any integer, and
\be
E^{(1)}={1\over 2R^2}\left\{ k^2+z^2
\left[\left({\omega^2\over\omega^3}+{\omega^3\over\omega^2}\right)
      \left(n_2+{1\over 2}\right)\left(n_3+{1\over 2}\right)
-{1\over 2}\right] -{1\over 4}\right\}.
\ee
Finer corrections to the spectrum may be evaluated by going over in
perturbation
theory. This matter not being particulary interesting for this model is
postponed to the next example which is physically more significative. The
remarkable fact  we learn from this example is that the
realization of the constraint, that is the particulary form of the confining
potential $V_C$ characterizes the spectrum and the effective dynamics of the
constrained quantum system. Such informations are completely lost
within a formal treatment of the constraint.

\section[]{Particle Constrained on a Sphere\\ (The Rigid Diatom)}

As a second example of a constrained quantomechanical system we consider the
motion of a particle on a sphere embedded in $R^3$.
To get some physical intuition on what
we are dealing with let us consider a diatom. Aside from effects connected to
the geometric phase \cite{MSW}, the effective Hamiltonian describing the
rotovibrational degrees of freedom of the molecule is written in the
adiabatic approximation as
\be
\H_{nuc}=-{\hbar^2\over 2\mu}\left(
{\partial^2\over {\partial \x}^2} +
{\partial^2\over {\partial \y}^2} +
{\partial^2\over {\partial \z}^2} \right) +
V_{BO}(|{\rb}|),
\label{Hnuc}
\ee
where ${\rb}=(\x,\y,\z)$ is the relative position of the nuclei,
$\mu$ the reduced mass of the system and $V_{BO}$ is the Born-Oppenheimer
potential. It is usual to assume that
\begin{description}
\item{---} $V_{BO}$ presents a deep minimum in correspondence of the
molecular equilibrium length $\r_0$,
\item{---} $V_{BO}$ depends only on the relative distance of the nuclei
$\r=|{\rb}|$ and not  on the orientation of the molecule in space.
\end{description}

That is, $V_{BO}$ behaves as a potential confining the motion from the
rotovibrational configuration space $R^3$ to the sphere of radius $\r_0$.
Hamiltonian (\ref{Hnuc}) describes therefore a constrained
quantomechanical system in the sense we specified before, crf. (\ref{Ha}).

In order to adapt coordinates we introduce the usual angles $\theta$ and $\phi$
parametrizing the sphere and the normal coordinate $y={1\over\r_0}
\sqrt{{I\omega\over\hbar}}(\r-\r_0)$. $I=\mu\r_0^2$ is the momentum of
inertia of the diatom and $\omega$ the frequency introduced by the
Born-Oppenheimer potential, $\omega=\sqrt{{2\over\mu}{\partial^2V_{BO}\over
{\partial\r}^2}|_{\r=\r_0}}$. The adimesional scale factor
$\epsilon=\hbar/I\omega$ appears naturally in the definition of $y$ once
the zero order energy $\hbar\omega$ is factorized from the Hamiltonian. From
most diatoms $\epsilon$ is a very small parameter, $\epsilon\simeq
10^{-2}-10^{-4}$, and, as our notation anticipates, gives a measure of the
rigidity of the molecule. The metric of $R^3$ in the adapted coordinates frame
reads
\be
G_{IJ} = {1\over \hbar\omega\epsilon}\pmatrix{
    \ds(1+\epsilon^{1/2}y)^2& 0 & 0 \cr
    0 & \ds(1+\epsilon^{1/2}y)^2\sin^2\theta & 0 \cr
    0 & 0 & \epsilon},
\label{Gsph}
\ee
while the Born-Oppenheimer potential writes as
\be
V_{BO}=\hbar\omega\left({1\over 2}y^2+{\hat a}y^3+{\hat b}y^4+....\right),
\ee
where ${\hat a}=a\hbar^{1/2}/\mu^{3/2}\omega^{5/2}$ and
      ${\hat b}=b\hbar/\mu^2\omega^3$, $a$ and $b$ being the usual
spectroscopical parameters.
The rigidity parameter ${\hbar/ I\omega}$ plays therefore the rule of the
parameter $\epsilon$ we introduced in section 3 for a generic
constraint\footnote{Note that the normal coordinate $y$ appears as already
rescaled.}. A comparison of (\ref{Gsph}) with equations (\ref{G}) and
(\ref{gamma})   allows to write down
immediately the induced metric and the second fundamental form on the sphere
\be
g_{\mu\nu} = {1\over \hbar\omega\epsilon}\pmatrix{
    \ds1& 0 \cr
    0 & \ds\sin^2\theta},
\ee
and
\be
\alpha_{\mu\nu} = -{\epsilon^{1/2} \over \hbar\omega}\pmatrix{
     1 & 0 \cr
     0 & \sin^2\theta}.
\ee
The codimension of the constraint surface being one, the normal fundamental
form vanishes identically. The embedding of the sphere in $R^3$ is standard and
the perturbative expansion $\H/\hbar\omega=H^{(0)}+\epsilon H^{(1)}+ ... +
{\hat a}y^3 + {\hat b}y^4 + ...$ is easily evaluated as
\be
\begin{array}{l}
\ds H^{(0)}={1\over 2}(-\partial^2_y+y^2), \\
\ds H^{(1)}=-{1\over 2}\left({1\over\sin\theta}{\partial_\theta}
\sin\theta {\partial_\theta} +{1\over\sin^2\theta}
{\partial^2_\phi}\right),\\
\ds ... , \\
\ds H^{(N/2)}=(-1)^{N}(N-1)y^{N-2} H^{(1)},\\
\ds ... \ \ .
\end{array}
\ee
The zero order Hamiltonian  $H^{(0)}$ takes into account the vibrational
motion of the diatom.
$H^{(1)}$ is the angular momentum operator describing the effective
rotational dynamics as that of a spherical top. The rest of the
perturbative expansion reproduces the Dunham expansion \cite{Dun} of the
nonrigid  rotator taking into account rotovibrational interactions.
The rotovibrational spectrum of the diatom
finds therefore a very natural interpretation in terms of constrained quantum
mechanics. Given up the classical idea of constraint ($\epsilon\rightarrow 0$)
the rotovibrational structure appears naturally as a consequence of the
physical structure of the constraint. The algorithm we present in this
paper gives an automatic way to compute rotovibrational interactions and may
result usefull in the analysis of rigid polyatomic molecules \cite{Mar}.
For the moment we
conclude by evaluating the spectrum of the particle constrained to the sphere.
Introduced creation/destruction operators relative to the normal coordinate the
computation results algebraic in nature and may be performed to an arbitrary
order in perturbation theory by means of computer algebraic manipulation. We
report here the expansion of the energy ${\cal E}_{n,l}$ to the third order in
perturbation theory
\be
\begin{array}{rl}
\ds{{\cal E}_{n,l}\over \hbar\omega}&\ds
=\left(n+{1\over 2}\right)+
 \epsilon   {1\over 2}\left[l(l+1)\right]+
 \epsilon^2 {3\over 2}\left[l(l+1)\right]\left(n+{1\over 2}\right)+\\
                                 &\ds
+ \epsilon^3 \left\{{15\over 8}\left[l(l+1)\right]+
                   {15\over 2}\left[l(l+1)\right]\left(n+{1\over 2}\right)^2-
                   {1 \over 2}\left[l(l+1)\right]^2
            \right\}+ \\
                                 &\ds
 +\left({3\over 2}{\hat b}-3\epsilon^{3/2}{\hat a}
       -{15\over 4}{\hat a}^2\right)\left(n+{1\over 2}\right)^2.
\end{array}
\label{Enl}
\ee
Replacing the values of $\epsilon$, ${\hat a}$ and ${\hat b}$
Eq.(\ref{Enl}) reproduces the
standard expression of rotovibrational energies of  diatoms \cite{Her}.

\section{Concluding Remarks}

The reduction of the motion of a quantomechanical system from its
configuration space to a submanifold is by no means  unique,
in the sense that it is impossible to perform this operation by completely
disregarding the motion in the directions normal to the constraint surface.
Quantum mechanics is a field theory, after all, and the wave function of the
system keeps on exploring the whole configuration space even if squeezed on the
constraint surface. When the system is in an eigenstate of the confining
potential we can obtain an effective Hamiltonian describing the dynamics on
the constraint surface. As is clearly illustrated by examples of
sections 5 and 6 this effective dynamics, described by ${\cal H}^{E^{(0)}}$,
depends both on the specific normal eigenstate and on the explicit form of
the confining potential. In any case the
eigenvalues of the effective Hamiltonian give only the first order corrections
to the spectrum of the system at finite $\epsilon$.
An accurate description requires also the
analysis of the interaction between the motion normal and along the
constraint surface. The perturbative expansion (\ref{Hexp}) we present in
this paper takes into account this effect.

 Performing the limit $\epsilon\rightarrow 0$ after subtracting the divergent
zero order energies of the system produces a well defined description of the
motion on the constraint surface. Nevertheless we consider this operation as
artificial, the physical nature of the constraint being in the small but finite
value of $\epsilon$ (cfr. the discussion of the diatom). The whole perturbative
expansion (\ref{Hexp}) is therefore necessary to characterize the dynamics of
the constrained system.

 Aside from its conceptual importance, the perturbative expansion
(\ref{Hexp}) may reveal of practical importance in the analysis of
electrons confined on arbitrary surfaces and wires as well as in the analysis
of polyatomic molecular spectra. The effective rotational dynamics of some
simple polyatomic molecules has already been considered in Ref.\cite{Mar}
demonstrating the physical relevance of the induced gauge structure and quantum
potentials (\ref{iA}), (\ref{iQ}) and (\ref{Qbar}). Our hope is that expansion
(\ref{Hexp}) may serve as an unifying tool in understanding the fine
structure spectra of rigid polyatomic molecules.

\section*{Aknoledgments}

I wish to warmly thank C.\ Destri and E.\ Onofri for useful discussions.

\end{document}